%% file: TrustRAG.tex
\documentclass[sigconf]{acmart}

\input{preamble/packages}
\input{preamble/definitions}

\input{preamble/authors}
\input{preamble/metadata}

\setcopyright{acmlicensed}
\copyrightyear{2018}
\acmYear{2018}
\acmDOI{XXXXXXX.XXXXXXX}

\acmISBN{978-1-4503-XXXX-X/18/06}

\begin{document}

\title{TrustRAG: An Information Assistant with Retrieval Augmented Generation}

\begin{abstract}
 \Ac{RAG} has emerged as a crucial technique for enhancing large models with real-time and domain-specific knowledge. While numerous improvements and open-source tools have been proposed to refine the \ac{RAG} framework for accuracy, relatively little attention has been given to improving the trustworthiness of generated results. To address this gap, we introduce TrustRAG, a novel framework that enhances \ac{RAG} from three perspectives: indexing, retrieval, and generation. Specifically, in the indexing stage, we propose a semantic-enhanced chunking strategy that incorporates hierarchical indexing to supplement each chunk with contextual information, ensuring semantic completeness. In the retrieval stage, we introduce a utility-based filtering mechanism to identify high-quality information, supporting answer generation while reducing input length. In the generation stage, we propose fine-grained citation enhancement, which detects opinion-bearing sentences in responses and infers citation relationships at the sentence-level, thereby improving citation accuracy. 
We open-source the TrustRAG framework and provide a demonstration studio designed for excerpt-based question answering tasks \footnote{https://huggingface.co/spaces/golaxy/TrustRAG}. Based on these, we aim to help researchers: 1) systematically enhancing the trustworthiness of \ac{RAG} systems and (2) developing their own \ac{RAG} systems with more reliable outputs.

\end{abstract}




\maketitle

\input{sections/1-introduction}

\input{sections/2-studio}

\input{sections/3-library}

\input{sections/4-usage}

\input{sections/5-future}

\bibliographystyle{ACM-Reference-Format}
\bibliography{sample-base}

\end{document}

%% file: preamble/packages.tex
\usepackage{multirow}
\usepackage{times}
\usepackage{tcolorbox} 
\usepackage[T1]{fontenc}
\usepackage{booktabs}
\usepackage[utf8]{inputenc}
\usepackage{makecell}
\usepackage{microtype}

\usepackage[inline]{enumitem}
\usepackage{amsmath}

\usepackage{graphicx}

\usepackage{acronym}
\usepackage{orcidlink} 

\usepackage[labelfont=bf]{caption}

\usepackage{listings}

%% file: preamble/definitions.tex
\AtBeginDocument{%
  \providecommand\BibTeX{{%
    \normalfont B\kern-0.5em{\scshape i\kern-0.25em b}\kern-0.8em\TeX}}}
    
\makeatletter
\g@addto@macro\normalsize{%
  \abovedisplayskip 3pt plus1pt
  \belowdisplayskip 3pt plus1pt
  \abovedisplayshortskip  0pt plus1pt
  \belowdisplayshortskip  0pt plus1pt
}
\makeatother

\acrodef{IS}{information seeking}
\acrodef{IR}{information retrieval}
\acrodef{LLM}{large language model}
\acrodef{NLP}{natural language processing} 
\acrodef{RAG}{Retrieval-Augmented Generation}

%% file: preamble/authors.tex
\author{
Yixing Fan$^{\dagger, \ddag}$ \orcidlink{0000-0003-4317-2702}, 
Qiang Yan$^{\dagger}$, 
Wenshan Wang$^{\dagger}$, \orcidlink{0009-0008-7390-3696}
Jiafeng Guo$^{\dagger, \ddag}$, 
Ruqing Zhang$^{\dagger, \ddag}$ 
and Xueqi Cheng$^{\dagger, \ddag}$
}
\affiliation{
  \institution{
  $^{\dagger}$Key Lab of Network Data Science and Technology, Institute of Computing Technology,\\ Chinese Academy of Sciences\\
  ${\ddag}$University of Chinese Academy of Sciences\\
  }
  \city{Beijing}
  \country{China} 
}
\email{{fanyixing, yanqiang, wangwenshan, guojiafeng, zhangruqing, cxq}@ict.ac.cn}

\renewcommand{\shortauthors}{Fan et al.}

%% file: preamble/metadata.tex
\setcopyright{rightsretained}
\copyrightyear{2025}
\acmYear{2025}
\acmDOI{XXXXXXX.XXXXXXX}

\acmConference[Conference acronym 'XX]{Make sure to enter the correct
  conference title from your rights confirmation emai}{June 03--05,
  2018}{Woodstock, NY}
\acmISBN{978-1-4503-XXXX-X/18/06}

%% file: sections/1-introduction.tex
\section{Introduction}
\Ac{IS} is one of the most important activities in our daily life and work. In the past few decades, search engines have become the main way people access information by locating relevant documents from the Web \cite{schutze2008introduction}. In recent years, the rapid development of \ac{LLM} has opened new opportunities to improve \ac{IS}, shifting from ranking relevant documents to producing reliable answers \cite{zhao2023survey}. 
However, generating answers directly with \ac{LLM} presents challenges, including missing real-time information, insufficient domain knowledge, and the risk of hallucinate claims, resulting in unreliable responses in real-world scenarios \cite{huang2024survey, tonmoy2024comprehensive, feng2023trends}.

To this end, \ac{RAG} has emerged as a promising solution by combining the strength of search systems and \acp{LLM} to improve the quality of results \cite{gao2023retrieval, zhao2024retrieval, li2024structrag}.
On one hand, \ac{RAG} leverages search to process external large corpus, enhancing access to real-time information. On the other hand, it utilizes \ac{LLM} to reason and generate text, improving the accuracy of the answer.
To better integrate the retriever and the generator, the researchers further developed various \ac{RAG} frameworks to improve the accuracy of answers, such as Self-RAG~\cite{asai2023self}, ActiveRAG~\cite{xu2024activerag}, CoRAG~\cite{wang2025chain}, etc.
In addition to improving accuracy, some studies also try to improve source attribution of the result, thus improving the reliability of results \cite{zhou2024trustworthiness, guan2025deeprag, friel2024ragbench}, such as InstructRAG~\cite{wei2024instructrag}, LongCite~\cite{zhang2024longcite}, SelfCite~\cite{chuang2025selfcite}, etc.
However, most of these works focus on improving specific aspects of the \ac{RAG} framework, while real-world applications require systematic enhancements across all components.
 
In addition to the above advances, researchers have created various open-source systems \cite{Liu_LlamaIndex_2022, guo2024lightrag, jin2024flashrag, zhang2024raglab} to support the development and practical application of \ac{RAG}.
For example, Langchain~\footnote{https://github.com/langchain-ai/langchain} is the most widely used \ac{RAG} framework, providing modular components for integrating \acp{LLM} with external data sources.
LLamaIndex~\footnote{https://github.com/run-llama/llama\_index} serves as a data framework designed to efficiently construct \ac{RAG} applications by streamlined data ingestion, indexing and querying processes.
LightRAG~\footnote{https://github.com/HKUDS/LightRAG} introduces a dual-level retrieval mechanism by incorporating the graph structure into the text indexing and retrieval. 
Nevertheless, these frameworks focus mainly on modularizing the process and simplifying the implementation of \ac{RAG} systems, further efforts are needed to improve the attribution of RAG-generated content. 

\begin{figure}[!t]
\centering
\includegraphics[scale=0.43]{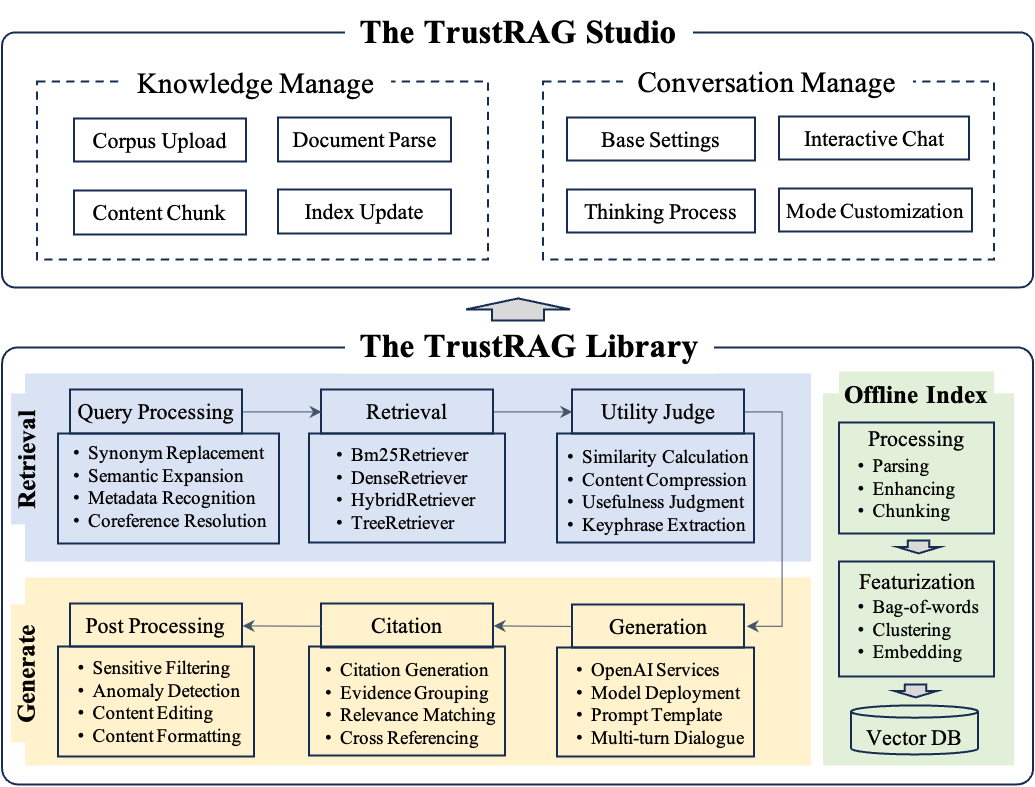}
\caption{An Overview of the System Architecture.}
\label{fig:architecture}
\end{figure}

In this demo, we introduce a novel \ac{RAG} system, named TrustRAG, to improve the accuracy and attribution of the result in a more comprehensive way. The overall architecture of the system consists of two main components: 
1) \textbf{the TrustRAG library}: an easy to use \ac{RAG} library which improves three major components of the RAG framework, including semantic-enhanced indexing, utility-enhanced retrieval, and attribution-enhanced generation; 
2) \textbf{the TrustRAG studio}: a user-friendly and interactive Web interface which enables users to browse, configure, experience, and create RAG applications.
Additionally, we provide an example application for \textit{Excerpt-Based Question Answering} (ExQA) built on TrustRAG. The ExQA is characterized by answers typically presented as a list, where each entry is derived from one or more documents and must be traceable to the original text. 
Our work makes the following key contributions:
\begin{itemize}
    \item \textbf{The TrustRAG Studio for No-Code \ac{RAG} Application Development}: The studio enables users to create their own \ac{RAG} applications without writing any code. It incorporates \textit{corpus management} and \textit{conversation management}--with corpus management, users can upload their own documents, configure chunking parameters,  and apply semantic-enhanced indexing, while conversation management enables them to adjust retrieval parameters, choose generation models, and engage in interactive Q\&A based on their corpus.
    \item \textbf{The TrustRAG Library for Low-Code \ac{RAG} Experimentation}: The library enables low-code experimentation with RAG by offering a comprehensive pipeline that covers indexing, retrieval, and generation, featuring over 20 modular components. Additionally, it incorporates semantic-enhanced indexing, utility-enhanced retrieval, and citation-enhanced generation, allowing users to flexibly combine these modules to build trustworthy RAG systems.
    \item \textbf{An Example Application for Excerpt-Based Question Answering}: The ExQA is a typical information seeking scenario that focuses on fine-grained information extraction from structured corpora. It aggregates and distills attribute-specific information from multiple sources to produce a final answer. The answers must be both clearly structured and fully traceable, with each entry linked to its corresponding text spans in the original documents. Common applications include Q\&A on legal documents, policies and regulations, and product manuals.
    \item \textbf{An open-source implementation} that lowers the barrier for researchers and developers to apply TrustRAG on the client side. We provide comprehensive documents to help users use TrustRAG to implement on-device \ac{RAG} system across different Web environments. The TrustRAG studio and library are publicly accessible at \url{https://huggingface.co/spaces/golaxy/TrustRAG} and \url{https://github.com/gomate-community/TrustRAG}.
\end{itemize}

%% file: sections/2-studio.tex
\section{System Overview}
The architecture of the system is shown in the Figure \ref{fig:architecture}. The system consists of two major components, namely the TrustRAG library and the TrustRAG studio. 

The library functions as the system's back-end, offering a comprehensive set of features for all stages of the \ac{RAG} pipeline. Its capabilities are structured into three modular components: the offline indexing module, the retrieval module, and the generation module.
First, the offline indexing module provide rich parsing functions for different kinds of files (e.g., PDF, Word, Excel, JSON) and converts chunked content into embeddings. 
Second, the retrieval module operates in three stages: query processing, retrieval, and utility assessment. 
Finally, the generation module also follows a three-stage process, comprising basic generation, citation integration, and post-processing.

The studio serves as the system's front-end, offering a user-friendly GUI built on the TrustRAG library. It features two main panels: \textbf{the knowledge manage panel} and \textbf{the conversation manage panel}. In the knowledge manage panel, users can upload their own documents, configure processing options, and select the indexing method. In the conversation manage panel, users can choose search method and the \ac{LLM} for each conversation. Additionally, the studio visualize the intermediate ``thinking'' process of the TrustRAG, including query understanding, document selection, answer reasoning, and sentence citation, to enhance reliability and transparency. 

 


%% file: sections/3-library.tex
\section{TrustRAG Library}
TrustRAG is a configurable and modular Retrieval-Augmented Generation (RAG) framework designed for "reliable input and trustworthy output." It consists of key components such as document parsing, text chunking, query optimization, retrieval ranking, content compression, model generation, and answer citation. This section highlights its innovations in semantic enhancement, usefulness enhancement, and citation enhancement.

\begin{figure}[!t]
\centering
\includegraphics[width=0.5\textwidth]{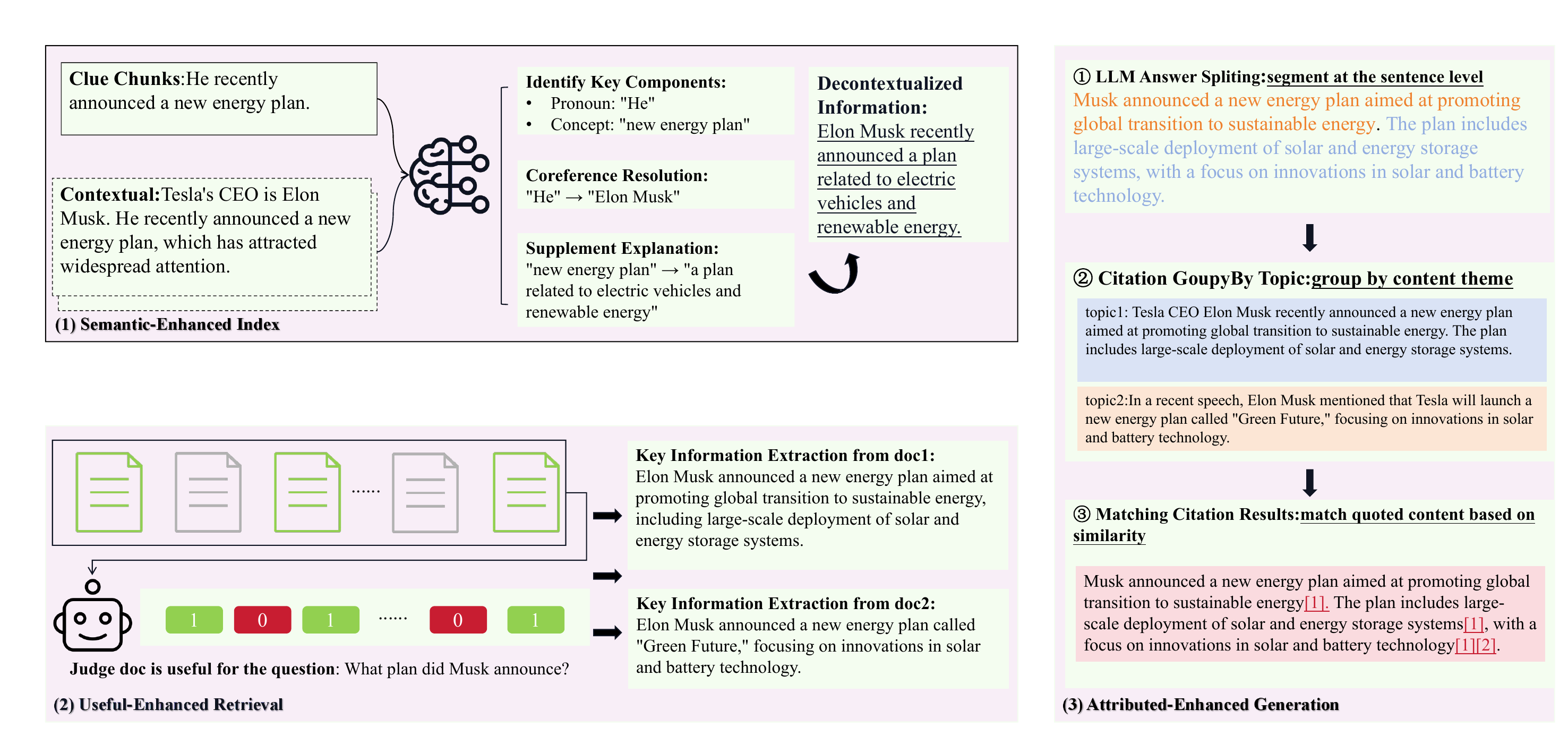}
\caption{An overview of the TrustRAG framework.}
\label{fig:trustrag_overview}
\end{figure}

\subsection{Semantic-Enhanced Indexing}
Existing text chunking methods, while efficient, often lead to significant semantic loss, particularly when handling long or complex documents~\cite{sarthi2024raptor}. Simple character-based or paragraph-based splitting can disrupt contextual coherence, making it difficult for downstream retrieval and generation tasks to fully utilize the semantic information embedded in the text. 

To address this issue, TrustRAG introduces the \textit{semantic-enhanced chunking} to improve the semantic integrity and coherence for each chunk. Specifically, we firstly take the \ac{LLM} to apply co-reference resolution for each document, which resolves ambiguities caused by pronouns or incomplete references. For instance, when a pronoun like "it" appears in a sentence, the system identifies its antecedent and restores the missing context, thereby enhancing the semantic completeness of the text. This process not only recovers lost semantic information but also provides more accurate contextual support for subsequent generation tasks.
Moreover, we standardize the time fields in the document by converting the relative time references into standardize date formats based on the document publication date. For example, if the document's publication date is ``2025-02-18'', terms like ``yesterday'' and ``last Friday'' will be converted to ``2025-02-17'' and ``2025-02-14'', respectively. 
This process not only recovers lost semantic information but also provides more accurate contextual support for subsequent generation tasks. The implementation of this feature can be found in \texttt{trustrag/modules/refiner/decontextualizer.py}.

Furthermore, TrustRAG supports advanced semantic segmentation techniques that dynamically identify semantic boundaries using embedding technologies and large language models (LLMs). Unlike static chunking methods, these techniques allow the system to adaptively split text based on its semantic structure, ensuring higher-quality chunks that preserve contextual coherence. The code is available in \texttt{trustrag/modules/chunks/semantic\_chunk.py}. These innovations improve the quality of text indexing, laying a solid foundation for reliable retrieval and generation.

\subsection{Utility-Enhanced Retrieval}
In conventional RAG systems, the relevance of retrieved documents is often determined solely by vector similarity. However, high similarity does not always translate to usefulness for the generation task. In some cases, even irrelevant documents may inadvertently improve system accuracy, highlighting the need for more intelligent mechanisms to evaluate the utility of retrieved results~\cite{cuconasu2024power}.
TrustRAG addresses this limitation by introducing two key innovations: \textbf{usefulness judgement} and \textbf{fine-grained evidence extraction}.
\begin{itemize}
    \item \textbf{Usefulness Judgement:} TrustRAG employs large language models (LLMs) as discriminators to assess the utility of retrieved documents. Through carefully designed prompts, the system evaluates the relevance of each document to the user's query and the generation task. This evaluation goes beyond surface-level similarity, incorporating deeper contextual understanding to ensure that only the most useful documents are selected. See \texttt{trustrag/modules/judger/llm\_judger.py}.
    
    \item \textbf{Fine-Grained Evidence Extraction:} After identifying useful documents, TrustRAG extracts the most relevant sentences through fine-grained evidence extraction. This process leverages model distillation techniques to reduce computational costs while maintaining high accuracy and relevance. By focusing on the most pertinent information, the system ensures that the generation task receives high-quality inputs. See \texttt{trustrag/modules/refiner/compressor.py}.
\end{itemize}
These enhancements enable TrustRAG to prioritize truly useful information, improving the overall quality and reliability of the retrieval process.

\subsection{Attribution-Enhanced Generation}
The credibility and traceability of generated answers are critical for trustworthiness in RAG systems. Traditional approaches rely heavily on direct reasoning by large models, which can be slow and prone to inaccuracies in citations. Additionally, fine-tuning models to improve citation accuracy may compromise their performance on other tasks, limiting practical applicability.

TrustRAG overcomes these challenges through two key innovations: \textbf{post-generation citation} and \textbf{citation grouping with cross-referencing}.
\begin{itemize}
    \item \textbf{Post-Generation Citation:} Instead of embedding citations during the generation process, TrustRAG matches the generated answers with retrieved reference materials afterward. This approach ensures higher citation accuracy while significantly accelerating the generation process. See \texttt{trustrag/mod
    ules/citation/match\_citation.py}.
    
    \item \textbf{Citation Grouping and Cross-Referencing:} To enhance traceability, TrustRAG organizes citations into logical groups, providing users with clearer reference sources. Furthermore, the system supports cross-referencing, allowing it to establish connections between different citations. This feature not only improves the clarity of references but also strengthens the credibility of the generated answers. See \texttt{trustrag/modules/
    citation/source\_citation.py}.
\end{itemize}
These innovations ensure that TrustRAG delivers both accurate and traceable answers, addressing key limitations of traditional RAG systems.

\begin{figure*}[!t]
\centering
\includegraphics[scale=0.6]{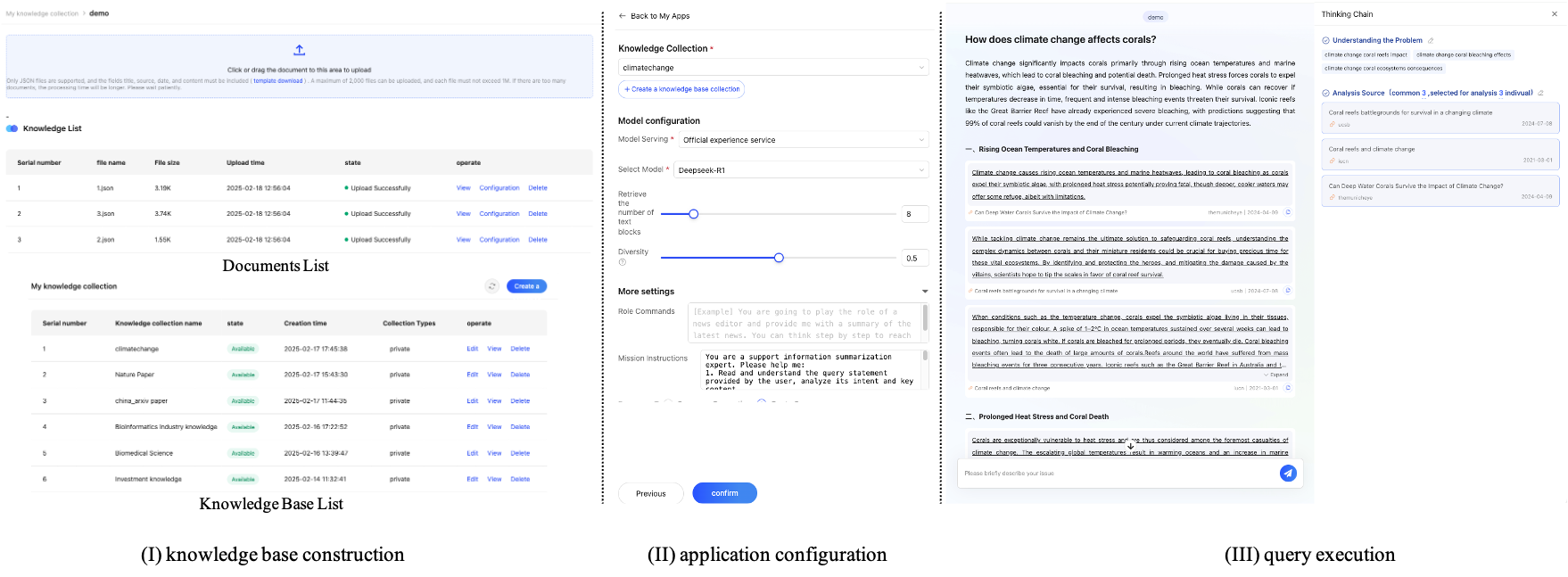}
\caption{Example usage of TrustRAG on Excerpt-based Questions}
\label{fig:demo}
\end{figure*}

\subsection{Additional Modules}
Beyond the three core enhancements, TrustRAG offers a rich set of modular functionalities, each designed to support specific aspects of the RAG pipeline:
\begin{itemize}
    \item \textbf{Document Parsing:} TrustRAG supports efficient parsing of multiple file formats, including PDF, Word, and HTML, with robust cross-language capabilities. This module ensures high parsing success rates and seamless integration with multilingual content. See \texttt{trustrag/modules/document} for implementation details.
    
    \item \textbf{Text Chunking:} From basic to advanced chunking methods, this module ensures semantic consistency and coherence in text segmentation. See \texttt{trustrag/modules/chunks}.
    
    \item \textbf{Query Optimization:} TrustRAG enhances query accuracy and efficiency through techniques such as query expansion, decomposition, disambiguation, and abstraction. These methods help refine user queries, improving the quality of retrieval results. See \texttt{trustrag/mod
    ules/rewriter}.
    
    \item \textbf{Retrieval Ranking:} By employing multi-path retrieval and fusion-based re-ranking, TrustRAG ensures high relevance and reliability in retrieval results. See \texttt{trustrag/modules/
    reranker}.
    
    \item \textbf{Content Compression:} This module extracts high-value information from large volumes of retrieved content using usefulness evaluation and semantic enhancement techniques. See \texttt{trustrag/modules/refiner}.
    
    \item \textbf{Model Generation:} TrustRAG supports flexible invocation of various large language models, providing efficient inference and indexing management capabilities. See \texttt{trustrag/
    modules/generator}.
\end{itemize}

Through these comprehensive innovations, TrustRAG significantly enhances the reliability and trustworthiness of RAG systems. Its modular and configurable design empowers users to tailor the framework to diverse application scenarios, delivering high-quality and trustworthy outputs across a wide range of use cases.

%% file: sections/4-usage.tex
\section{System Usage}
The typical use case for \ac{RAG} systems involves the summarization of retrieved information to produce concise, generalized content. Numerous studies have significantly advanced the fluency of generated language, the comprehensiveness of the covered information, and the ability to extract core insights. However, there is a notable gap in research concerning the verification of the consistency between the generated output and the original source text.

TrustRAG addresses the ExQA task, with the primary goal of generating content that is both accurate and traceable to the original documents, while clearly indicating the sources of the information. This task scenario emphasizes strict adherence to the constraints of the original reference materials during the generation process, ensuring the credibility of the resulting answer. 

To illustrate the application of the TrustRAG, we present a case study focused on news related to climate change. As depicted in Figure \ref{fig:demo}, the system follows three key steps: (I) constructing a knowledge base and uploading relevant documents, (II) configuring the question-answering application, which involves selecting the appropriate knowledge base, choosing the suitable generation model, and specifying the desired output format, and (III) executing the question-answering task within the generated application instance.

As shown in step (III) of Figure \ref{fig:demo}, TrustRAG presents system outputs in a clear, concise, and structured format:
\begin{itemize}
    \item The answer display begins with a brief summary of a few sentences, providing a high-level response to the user's query. For example, in response to the query "How does climate change affects corals?", concise summary is presented at the top. This summary allows the user to quickly understand the overall output of the system.. 
    \item Key aspects of the response are presented separately, each organized under a bolded subheading to facilitate efficient navigation. In this case, the answer is structured into three main aspects: 1. Rising Ocean Temperatures and Coral Bleaching, 2. Prolonged Heat Stress and Coral Death, 3. Impact on Iconic Coral Reefs.
    \item Beneath each subheading, evidence is directly sourced from the retrieved documents, with the system clearly listing each source, including the title of the article, author(s), publication date, and a link to the original material. 
    \item A right-side panel provides detailed information regarding the system's reasoning process, illustrating how the system interprets the user's query intent and the rationale behind selecting relevant information from the knowledge base.
\end{itemize}

%% file: sections/5-future.tex
\section{Conclusion}
In this demo, we introduce a novel TrustRAG system for risk-aware information seeking scenarios. Users can build their own \ac{RAG} applications with private corpus, study the \ac{RAG} components within the library, and experiment with \ac{RAG} library with customized modules. 
We will showcase the TrustRAG system through the following aspects: (1) we will use a poster to give an overview of the system and briefly show the pipeline of the framework. (2) We will demonstrate how to use the system to create \ac{RAG} application with a private corpus. (3) We will share insights on system's strengths and limitations, along with potential future enhancements.

%% file: TrustRAG.bbl

\begin{thebibliography}{23}


\ifx \showCODEN    \undefined \def \showCODEN     #1{\unskip}     \fi
\ifx \showDOI      \undefined \def \showDOI       #1{#1}\fi
\ifx \showISBNx    \undefined \def \showISBNx     #1{\unskip}     \fi
\ifx \showISBNxiii \undefined \def \showISBNxiii  #1{\unskip}     \fi
\ifx \showISSN     \undefined \def \showISSN      #1{\unskip}     \fi
\ifx \showLCCN     \undefined \def \showLCCN      #1{\unskip}     \fi
\ifx \shownote     \undefined \def \shownote      #1{#1}          \fi
\ifx \showarticletitle \undefined \def \showarticletitle #1{#1}   \fi
\ifx \showURL      \undefined \def \showURL       {\relax}        \fi
\providecommand\bibfield[2]{#2}
\providecommand\bibinfo[2]{#2}
\providecommand\natexlab[1]{#1}
\providecommand\showeprint[2][]{arXiv:#2}

\bibitem[Asai et~al\mbox{.}(2023)]%
        {asai2023self}
\bibfield{author}{\bibinfo{person}{Akari Asai}, \bibinfo{person}{Zeqiu Wu},
  \bibinfo{person}{Yizhong Wang}, \bibinfo{person}{Avirup Sil}, {and}
  \bibinfo{person}{Hannaneh Hajishirzi}.} \bibinfo{year}{2023}\natexlab{}.
\newblock \showarticletitle{Self-rag: Learning to retrieve, generate, and
  critique through self-reflection}.
\newblock \bibinfo{journal}{\emph{arXiv preprint arXiv:2310.11511}}
  (\bibinfo{year}{2023}).
\newblock


\bibitem[Chuang et~al\mbox{.}(2025)]%
        {chuang2025selfcite}
\bibfield{author}{\bibinfo{person}{Yung-Sung Chuang}, \bibinfo{person}{Benjamin
  Cohen-Wang}, \bibinfo{person}{Shannon~Zejiang Shen},
  \bibinfo{person}{Zhaofeng Wu}, \bibinfo{person}{Hu Xu},
  \bibinfo{person}{Xi~Victoria Lin}, \bibinfo{person}{James Glass},
  \bibinfo{person}{Shang-Wen Li}, {and} \bibinfo{person}{Wen-tau Yih}.}
  \bibinfo{year}{2025}\natexlab{}.
\newblock \showarticletitle{SelfCite: Self-Supervised Alignment for Context
  Attribution in Large Language Models}.
\newblock \bibinfo{journal}{\emph{arXiv preprint arXiv:2502.09604}}
  (\bibinfo{year}{2025}).
\newblock


\bibitem[Cuconasu et~al\mbox{.}(2024)]%
        {cuconasu2024power}
\bibfield{author}{\bibinfo{person}{Florin Cuconasu}, \bibinfo{person}{Giovanni
  Trappolini}, \bibinfo{person}{Federico Siciliano}, \bibinfo{person}{Simone
  Filice}, \bibinfo{person}{Cesare Campagnano}, \bibinfo{person}{Yoelle
  Maarek}, \bibinfo{person}{Nicola Tonellotto}, {and} \bibinfo{person}{Fabrizio
  Silvestri}.} \bibinfo{year}{2024}\natexlab{}.
\newblock \showarticletitle{The power of noise: Redefining retrieval for rag
  systems}. In \bibinfo{booktitle}{\emph{Proceedings of the 47th International
  ACM SIGIR Conference on Research and Development in Information Retrieval}}.
  \bibinfo{pages}{719--729}.
\newblock


\bibitem[Feng et~al\mbox{.}(2023)]%
        {feng2023trends}
\bibfield{author}{\bibinfo{person}{Zhangyin Feng}, \bibinfo{person}{Weitao Ma},
  \bibinfo{person}{Weijiang Yu}, \bibinfo{person}{Lei Huang},
  \bibinfo{person}{Haotian Wang}, \bibinfo{person}{Qianglong Chen},
  \bibinfo{person}{Weihua Peng}, \bibinfo{person}{Xiaocheng Feng},
  \bibinfo{person}{Bing Qin}, {et~al\mbox{.}}} \bibinfo{year}{2023}\natexlab{}.
\newblock \showarticletitle{Trends in integration of knowledge and large
  language models: A survey and taxonomy of methods, benchmarks, and
  applications}.
\newblock \bibinfo{journal}{\emph{arXiv preprint arXiv:2311.05876}}
  (\bibinfo{year}{2023}).
\newblock


\bibitem[Friel et~al\mbox{.}(2024)]%
        {friel2024ragbench}
\bibfield{author}{\bibinfo{person}{Robert Friel}, \bibinfo{person}{Masha
  Belyi}, {and} \bibinfo{person}{Atindriyo Sanyal}.}
  \bibinfo{year}{2024}\natexlab{}.
\newblock \showarticletitle{Ragbench: Explainable benchmark for
  retrieval-augmented generation systems}.
\newblock \bibinfo{journal}{\emph{arXiv preprint arXiv:2407.11005}}
  (\bibinfo{year}{2024}).
\newblock


\bibitem[Gao et~al\mbox{.}(2023)]%
        {gao2023retrieval}
\bibfield{author}{\bibinfo{person}{Yunfan Gao}, \bibinfo{person}{Yun Xiong},
  \bibinfo{person}{Xinyu Gao}, \bibinfo{person}{Kangxiang Jia},
  \bibinfo{person}{Jinliu Pan}, \bibinfo{person}{Yuxi Bi}, \bibinfo{person}{Yi
  Dai}, \bibinfo{person}{Jiawei Sun}, {and} \bibinfo{person}{Haofen Wang}.}
  \bibinfo{year}{2023}\natexlab{}.
\newblock \showarticletitle{Retrieval-augmented generation for large language
  models: A survey}.
\newblock \bibinfo{journal}{\emph{arXiv preprint arXiv:2312.10997}}
  (\bibinfo{year}{2023}).
\newblock


\bibitem[Guan et~al\mbox{.}(2025)]%
        {guan2025deeprag}
\bibfield{author}{\bibinfo{person}{Xinyan Guan}, \bibinfo{person}{Jiali Zeng},
  \bibinfo{person}{Fandong Meng}, \bibinfo{person}{Chunlei Xin},
  \bibinfo{person}{Yaojie Lu}, \bibinfo{person}{Hongyu Lin},
  \bibinfo{person}{Xianpei Han}, \bibinfo{person}{Le Sun}, {and}
  \bibinfo{person}{Jie Zhou}.} \bibinfo{year}{2025}\natexlab{}.
\newblock \showarticletitle{DeepRAG: Thinking to Retrieval Step by Step for
  Large Language Models}.
\newblock \bibinfo{journal}{\emph{arXiv preprint arXiv:2502.01142}}
  (\bibinfo{year}{2025}).
\newblock


\bibitem[Guo et~al\mbox{.}(2024)]%
        {guo2024lightrag}
\bibfield{author}{\bibinfo{person}{Zirui Guo}, \bibinfo{person}{Lianghao Xia},
  \bibinfo{person}{Yanhua Yu}, \bibinfo{person}{Tu Ao}, {and}
  \bibinfo{person}{Chao Huang}.} \bibinfo{year}{2024}\natexlab{}.
\newblock \showarticletitle{LightRAG: Simple and Fast Retrieval-Augmented
  Generation}.
\newblock  (\bibinfo{year}{2024}).
\newblock
\showeprint[arxiv]{2410.05779}~[cs.IR]


\bibitem[Huang et~al\mbox{.}(2024)]%
        {huang2024survey}
\bibfield{author}{\bibinfo{person}{Lei Huang}, \bibinfo{person}{Weijiang Yu},
  \bibinfo{person}{Weitao Ma}, \bibinfo{person}{Weihong Zhong},
  \bibinfo{person}{Zhangyin Feng}, \bibinfo{person}{Haotian Wang},
  \bibinfo{person}{Qianglong Chen}, \bibinfo{person}{Weihua Peng},
  \bibinfo{person}{Xiaocheng Feng}, \bibinfo{person}{Bing Qin},
  {et~al\mbox{.}}} \bibinfo{year}{2024}\natexlab{}.
\newblock \showarticletitle{A survey on hallucination in large language models:
  Principles, taxonomy, challenges, and open questions}.
\newblock \bibinfo{journal}{\emph{ACM Transactions on Information Systems}}
  (\bibinfo{year}{2024}).
\newblock


\bibitem[Jin et~al\mbox{.}(2024)]%
        {jin2024flashrag}
\bibfield{author}{\bibinfo{person}{Jiajie Jin}, \bibinfo{person}{Yutao Zhu},
  \bibinfo{person}{Xinyu Yang}, \bibinfo{person}{Chenghao Zhang}, {and}
  \bibinfo{person}{Zhicheng Dou}.} \bibinfo{year}{2024}\natexlab{}.
\newblock \showarticletitle{FlashRAG: A Modular Toolkit for Efficient
  Retrieval-Augmented Generation Research}.
\newblock \bibinfo{journal}{\emph{arXiv preprint arXiv:2405.13576}}
  (\bibinfo{year}{2024}).
\newblock


\bibitem[Li et~al\mbox{.}(2024)]%
        {li2024structrag}
\bibfield{author}{\bibinfo{person}{Zhuoqun Li}, \bibinfo{person}{Xuanang Chen},
  \bibinfo{person}{Haiyang Yu}, \bibinfo{person}{Hongyu Lin},
  \bibinfo{person}{Yaojie Lu}, \bibinfo{person}{Qiaoyu Tang},
  \bibinfo{person}{Fei Huang}, \bibinfo{person}{Xianpei Han},
  \bibinfo{person}{Le Sun}, {and} \bibinfo{person}{Yongbin Li}.}
  \bibinfo{year}{2024}\natexlab{}.
\newblock \showarticletitle{Structrag: Boosting knowledge intensive reasoning
  of llms via inference-time hybrid information structurization}.
\newblock \bibinfo{journal}{\emph{arXiv preprint arXiv:2410.08815}}
  (\bibinfo{year}{2024}).
\newblock


\bibitem[Liu(2022)]%
        {Liu_LlamaIndex_2022}
\bibfield{author}{\bibinfo{person}{Jerry Liu}.}
  \bibinfo{year}{2022}\natexlab{}.
\newblock \bibinfo{booktitle}{\emph{{LlamaIndex}}}.
\newblock
\urldef\tempurl%
\url{https://doi.org/10.5281/zenodo.1234}
\showDOI{\tempurl}


\bibitem[Sarthi et~al\mbox{.}(2024)]%
        {sarthi2024raptor}
\bibfield{author}{\bibinfo{person}{Parth Sarthi}, \bibinfo{person}{Salman
  Abdullah}, \bibinfo{person}{Aditi Tuli}, \bibinfo{person}{Shubh Khanna},
  \bibinfo{person}{Anna Goldie}, {and} \bibinfo{person}{Christopher~D
  Manning}.} \bibinfo{year}{2024}\natexlab{}.
\newblock \showarticletitle{Raptor: Recursive abstractive processing for
  tree-organized retrieval}.
\newblock \bibinfo{journal}{\emph{arXiv preprint arXiv:2401.18059}}
  (\bibinfo{year}{2024}).
\newblock


\bibitem[Sch{\"u}tze et~al\mbox{.}(2008)]%
        {schutze2008introduction}
\bibfield{author}{\bibinfo{person}{Hinrich Sch{\"u}tze},
  \bibinfo{person}{Christopher~D Manning}, {and} \bibinfo{person}{Prabhakar
  Raghavan}.} \bibinfo{year}{2008}\natexlab{}.
\newblock \bibinfo{booktitle}{\emph{Introduction to information retrieval}}.
  Vol.~\bibinfo{volume}{39}.
\newblock \bibinfo{publisher}{Cambridge University Press Cambridge}.
\newblock


\bibitem[Tonmoy et~al\mbox{.}(2024)]%
        {tonmoy2024comprehensive}
\bibfield{author}{\bibinfo{person}{SM Tonmoy}, \bibinfo{person}{SM Zaman},
  \bibinfo{person}{Vinija Jain}, \bibinfo{person}{Anku Rani},
  \bibinfo{person}{Vipula Rawte}, \bibinfo{person}{Aman Chadha}, {and}
  \bibinfo{person}{Amitava Das}.} \bibinfo{year}{2024}\natexlab{}.
\newblock \showarticletitle{A comprehensive survey of hallucination mitigation
  techniques in large language models}.
\newblock \bibinfo{journal}{\emph{arXiv preprint arXiv:2401.01313}}
  (\bibinfo{year}{2024}).
\newblock


\bibitem[Wang et~al\mbox{.}(2025)]%
        {wang2025chain}
\bibfield{author}{\bibinfo{person}{Liang Wang}, \bibinfo{person}{Haonan Chen},
  \bibinfo{person}{Nan Yang}, \bibinfo{person}{Xiaolong Huang},
  \bibinfo{person}{Zhicheng Dou}, {and} \bibinfo{person}{Furu Wei}.}
  \bibinfo{year}{2025}\natexlab{}.
\newblock \showarticletitle{Chain-of-Retrieval Augmented Generation}.
\newblock \bibinfo{journal}{\emph{arXiv preprint arXiv:2501.14342}}
  (\bibinfo{year}{2025}).
\newblock


\bibitem[Wei et~al\mbox{.}(2024)]%
        {wei2024instructrag}
\bibfield{author}{\bibinfo{person}{Zhepei Wei}, \bibinfo{person}{Wei-Lin Chen},
  {and} \bibinfo{person}{Yu Meng}.} \bibinfo{year}{2024}\natexlab{}.
\newblock \showarticletitle{InstructRAG: Instructing Retrieval Augmented
  Generation via Self-Synthesized Rationales}. In
  \bibinfo{booktitle}{\emph{Adaptive Foundation Models: Evolving AI for
  Personalized and Efficient Learning}}.
\newblock


\bibitem[Xu et~al\mbox{.}(2024)]%
        {xu2024activerag}
\bibfield{author}{\bibinfo{person}{Zhipeng Xu}, \bibinfo{person}{Zhenghao Liu},
  \bibinfo{person}{Yibin Liu}, \bibinfo{person}{Chenyan Xiong},
  \bibinfo{person}{Yukun Yan}, \bibinfo{person}{Shuo Wang},
  \bibinfo{person}{Shi Yu}, \bibinfo{person}{Zhiyuan Liu}, {and}
  \bibinfo{person}{Ge Yu}.} \bibinfo{year}{2024}\natexlab{}.
\newblock \showarticletitle{Activerag: Revealing the treasures of knowledge via
  active learning}.
\newblock \bibinfo{journal}{\emph{arXiv preprint arXiv:2402.13547}}
  (\bibinfo{year}{2024}).
\newblock


\bibitem[Zhang et~al\mbox{.}(2024a)]%
        {zhang2024longcite}
\bibfield{author}{\bibinfo{person}{Jiajie Zhang}, \bibinfo{person}{Yushi Bai},
  \bibinfo{person}{Xin Lv}, \bibinfo{person}{Wanjun Gu},
  \bibinfo{person}{Danqing Liu}, \bibinfo{person}{Minhao Zou},
  \bibinfo{person}{Shulin Cao}, \bibinfo{person}{Lei Hou},
  \bibinfo{person}{Yuxiao Dong}, \bibinfo{person}{Ling Feng}, {et~al\mbox{.}}}
  \bibinfo{year}{2024}\natexlab{a}.
\newblock \showarticletitle{Longcite: Enabling llms to generate fine-grained
  citations in long-context qa}.
\newblock \bibinfo{journal}{\emph{arXiv preprint arXiv:2409.02897}}
  (\bibinfo{year}{2024}).
\newblock


\bibitem[Zhang et~al\mbox{.}(2024b)]%
        {zhang2024raglab}
\bibfield{author}{\bibinfo{person}{Xuanwang Zhang}, \bibinfo{person}{Yunze
  Song}, \bibinfo{person}{Yidong Wang}, \bibinfo{person}{Shuyun Tang},
  \bibinfo{person}{Xinfeng Li}, \bibinfo{person}{Zhengran Zeng},
  \bibinfo{person}{Zhen Wu}, \bibinfo{person}{Wei Ye}, \bibinfo{person}{Wenyuan
  Xu}, \bibinfo{person}{Yue Zhang}, {et~al\mbox{.}}}
  \bibinfo{year}{2024}\natexlab{b}.
\newblock \showarticletitle{Raglab: A modular and research-oriented unified
  framework for retrieval-augmented generation}.
\newblock \bibinfo{journal}{\emph{arXiv preprint arXiv:2408.11381}}
  (\bibinfo{year}{2024}).
\newblock


\bibitem[Zhao et~al\mbox{.}(2024)]%
        {zhao2024retrieval}
\bibfield{author}{\bibinfo{person}{Penghao Zhao}, \bibinfo{person}{Hailin
  Zhang}, \bibinfo{person}{Qinhan Yu}, \bibinfo{person}{Zhengren Wang},
  \bibinfo{person}{Yunteng Geng}, \bibinfo{person}{Fangcheng Fu},
  \bibinfo{person}{Ling Yang}, \bibinfo{person}{Wentao Zhang}, {and}
  \bibinfo{person}{Bin Cui}.} \bibinfo{year}{2024}\natexlab{}.
\newblock \showarticletitle{Retrieval-augmented generation for ai-generated
  content: A survey}.
\newblock \bibinfo{journal}{\emph{arXiv preprint arXiv:2402.19473}}
  (\bibinfo{year}{2024}).
\newblock


\bibitem[Zhao et~al\mbox{.}(2023)]%
        {zhao2023survey}
\bibfield{author}{\bibinfo{person}{Wayne~Xin Zhao}, \bibinfo{person}{Kun Zhou},
  \bibinfo{person}{Junyi Li}, \bibinfo{person}{Tianyi Tang},
  \bibinfo{person}{Xiaolei Wang}, \bibinfo{person}{Yupeng Hou},
  \bibinfo{person}{Yingqian Min}, \bibinfo{person}{Beichen Zhang},
  \bibinfo{person}{Junjie Zhang}, \bibinfo{person}{Zican Dong},
  {et~al\mbox{.}}} \bibinfo{year}{2023}\natexlab{}.
\newblock \showarticletitle{A survey of large language models}.
\newblock \bibinfo{journal}{\emph{arXiv preprint arXiv:2303.18223}}
  (\bibinfo{year}{2023}).
\newblock


\bibitem[Zhou et~al\mbox{.}(2024)]%
        {zhou2024trustworthiness}
\bibfield{author}{\bibinfo{person}{Yujia Zhou}, \bibinfo{person}{Yan Liu},
  \bibinfo{person}{Xiaoxi Li}, \bibinfo{person}{Jiajie Jin},
  \bibinfo{person}{Hongjin Qian}, \bibinfo{person}{Zheng Liu},
  \bibinfo{person}{Chaozhuo Li}, \bibinfo{person}{Zhicheng Dou},
  \bibinfo{person}{Tsung-Yi Ho}, {and} \bibinfo{person}{Philip~S Yu}.}
  \bibinfo{year}{2024}\natexlab{}.
\newblock \showarticletitle{Trustworthiness in retrieval-augmented generation
  systems: A survey}.
\newblock \bibinfo{journal}{\emph{arXiv preprint arXiv:2409.10102}}
  (\bibinfo{year}{2024}).
\newblock


\end{thebibliography}
